\documentstyle{article}

\begin{document}

\title{Gravitational Meissner Effect}
\author{R.P. Lano \\
Department of Physics and Astronomy\\
The University of Iowa\\
Iowa City, IA, 52242}
\date{March 10, 1996}
\maketitle

\begin{abstract}
The gravitational analogue of the electromagnetic Meissner effect is
investigated. Starting from the post-Newtonian approximation to general
relativity we arrive at gravitational London equations, predicting a
gravitational Meissner effect. Applied to neutron stars we arrive at a
London penetration depth of 12km, which is about the size of a neutron star.

\vspace{12pt}

\noindent PACS number(s): 04.60, 74.25.H, 97.60.J, 04.70
\end{abstract}

\begin{center}
\section{Introduction}
\end{center}

There is strong evidence that neutron stars are superfluid and maybe
superconducting \cite{Pin-92}, indicating that macroscopic quantum effects
play a significant role in neutron stars. In addition neutron stars are very
strongly gravitating objects. We therefore would like to investigate the
possibility of macroscopic quantum gravity effect in the form of a
gravitational Meissner effect.

In the post-Newtonian approximation Einstein's theory of general relativity,
gravity looks like a U(1) theory and is very similar to the electromagnetic
Maxwell theory \cite{Tho-88}. This allows us to follow closely the approach
of London \cite{Lon-48, Lon-50} in deriving gravitational London equations,
predicting a gravitational Meissner effect. Not knowing the underlying
microscopic theory of quantum gravity we have to resort to this
semi-classical approach. We find the gravitational London penetration depth
to be about $12$ km for a typical neutron star of radius $15$ km.
Furthermore, as the density of the neutron star increases, the London
penetration depth decreases, suggesting that quantum gravity aspects may
become significant as the neutron star approaches collapse.

Macroscopic quantum aspects are best described by a BCS type theory\cite
{Sch-64}. However, since BCS needs a quantum theory of the underlying
interaction, and since we do not have a microscopic theory of quantum
gravity at this moment, a gravitational BCS type theory is not within our
reach. Despite this, we can certainly make a plausible argument for what
happens at the microscopic scale. Inspired from the Cooper-pair forming
electrons in the BCS theory of superconductivity, the Interacting Boson
Model (IBM) \cite{Iac-93, Muk-89} assumes nucleons to form pairs inside
nuclei. These pairs are bosons and by only looking at the interaction of
those bosons among themselves, one can explain a wealth of nuclear data to a
high degree of accuracy. Neutron stars consist mostly of neutrons and some
protons and for low enough temperatures ($<10^9$K) they are expected to pair
up. Experimental evidence for the pairing comes from the theory of cooling
of neutron stars, where the reduction of the specific heat due to extensive
pairing is essential \cite{Pag-94}. In addition, bose-condensation can occur
and superfluidity as well as superconductivity are consequences. Evidence
for superfluidity can be found in the glitches in the timing history of
pulsars \cite{Pin-92, Sau-89}. Therefore, although we do not have a
microscopic theory of quantum gravity, we have enough evidence justifying a
phenomenological investigation of macroscopic quantum gravity effects
following London's approach.

Symmetry breaking is a very convenient way to look at macroscopic quantum
effects. To break a symmetry one needs a bose-condensate. What causes the
bose-condensation and which symmetry is broken are two entirely unrelated
problems. Any mechanism leading to an attraction between the fermions at the
Fermi surface is equally well suited for producing the bose-condensate. For
example, in the standard BCS theory of superconductivity, the attraction is
caused by the phonon interaction, the symmetry broken is the electromagnetic
U(1) symmetry \cite{Ryd-85}. In the theory of superfluid helium 3 \cite
{Vol-90}, the attraction is caused by van der Waals forces, the symmetries
broken are translational \cite{Mer-78} and rotational in spin and orbital
space as well as gauge symmetry. In the theory of the Interacting Boson
Model, the attraction is caused by the strong nuclear force, the symmetry
broken is related to angular momentum and Runge-Lenz vector \cite{Iac-93}.
In the theory of superconductivity of protons inside neutron stars, the
attraction is caused by the strong nuclear force, the symmetry broken is the
electromagnetic U(1) symmetry. Therefore, we see no reason why to exclude
the scenario, where the attraction is caused by the strong nuclear force and
the symmetry broken is gravitational.

This work is also an attempt to probe certain aspects of quantum gravity.
The microscopic approach to quantum gravity has been poised with serious
problems, and no satisfactory theory has emerged as of yet. The macroscopic
approach seems much more tractable, and maybe following the theory of
superconductivity from London to Ginzburg and Landau to Bardeen, Cooper and
Schrieffer will give us another route to learn more about quantum gravity.

\begin{center}
\section{Gravitational Maxwell Equations}
\end{center}

Let us review the post-Newtonian approximation \cite{Wei-72, Mis-73}. For
weak gravitational fields and low velocities, i.e., 
\[
U/c^2\sim \left( v/c\right) ^4\sim \left( p/\rho c^2\right) ^2\sim \left(
\Pi /c^2\right) ^2\ll 1, 
\]
Einstein's field equation can be written in a form resembling Maxwell's
equations of electrodynamics. We are mostly interested in magnetic-type
gravity, and therefore we shall use the truncated and rewritten version of
the parametrized-post-Newtonian (PPN) formalism given by Braginsky et. al. 
\cite{Bar-77}. Scalar and vector potentials are defined as

\begin{equation}
\phi =-\left( U+2\Psi \right) \ \qquad \rm{and\qquad \ }A_j=-\frac
72V_j-\frac 12W_j,  \label{E2.1}
\end{equation}
where $U,\Psi ,V_j,$ and $W_j$ are defined in chapter 39 of Misner, Thorne
and Wheeler \cite{Mis-73}. This allows us to define a 'gravitoelectric'
field ${\bf g}$ and a 'gravitomagnetic' field ${\bf H}$ through

\begin{equation}
{\bf g=-\nabla }\phi -\frac 1c\frac{\partial {\bf A}}{\partial t}\ \rm{
\qquad and\qquad \ }{\bf H=\nabla }\times {\bf A.}  \label{E2.2}
\end{equation}

In this weak field, slow motion expansion of general relativity, ${\bf g}$
contains mostly first order corrections to flat space-time, and ${\bf H}$
contains second order corrections. For the solar system, ${\bf g}$ is just
the normal Newtonian gravitational acceleration, whereas ${\bf H}$ is
related to angular momentum interactions and effects due to ${\bf H}$ are
about $10^{12}$ times smaller than those due to ${\bf g}$.

In the above approximation, Einstein's field equations can be rewritten in a
form very much resembling that of Maxwell's equations

\begin{equation}
{\bf \nabla \cdot g=-}4\pi G\rho _0+{\cal O}\left( c^{-2}\right) \rm{%
,\qquad \ }{\bf \nabla \times g=}-\frac 1c\frac{\partial {\bf H}}{\partial t}
\label{E2.3}
\end{equation}

\begin{equation}
{\bf \nabla \cdot H}=0\rm{,\qquad \ }{\bf \nabla \times H}=4\left( -4\pi
G\rho _0\frac{{\bf v}}c+\frac 1c\frac{\partial {\bf g}}{\partial t}\right) ,
\label{E2.4}
\end{equation}
where $\rho _0$ is the density of rest mass in the local rest frame of the
matter and ${\bf v}$ is the ordinary velocity of the rest mass relative to
the PPN coordinate frame. The specific internal energy $\Pi $ and the
pressure $p$ have been neglected since they only contribute ${\cal O}\left(
c^{-2}\right) $ effects to the electric field.

The equation of motion of an uncharged particle is identical to the
electromagnetic Lorentz force law \cite{Tho-88, Bar-77},

\begin{equation}
\frac{d{\bf v}}{dt}={\bf g}+\frac 1c\,{\bf v\times H}+{\cal O}\left(
c^{-2}\right)  \label{E2.5}
\end{equation}

\begin{center}
\section{Gravitational London Equations}
\end{center}

With these gravitational Maxwell equations we can derive the equivalent of
London's equations for gravity. The derivation is completely analogous to
the electromagnetic case \cite{Lon-50, Sch-64, Ash-76, deG-66}. The Lorentz
force law tells us that given a gravitational electric field there will be a
resultant acceleration. Multiplying by $\rho _0$, the matter density, and
defining the matter current ${\bf j}=\rho _0{\bf v}$, we arrive at

\begin{equation}
\frac{d{\bf j}}{dt}=\rho _0{\bf g.}  \label{E3.3}
\end{equation}
This is sometimes referred to as the first London equation. Now substitute
this into the right equation of (\ref{E2.3}), the equivalent of Faraday's
law, we obtain

\begin{equation}
\frac \partial {\partial t}\left( \frac 1{\rho _0}{\bf \nabla \times j}%
+\frac 1c{\bf H}\right) =0.  \label{E3.4}
\end{equation}

One possible solution to the above equation is obtained by integrating and
setting the integration constant to zero. This leads to the second (and most
famous) London equation

\begin{equation}
{\bf \nabla \times j}=-\frac{\rho _0}c{\bf H,}  \label{E3.5}
\end{equation}
which predicts the Meissner-Ochsenfeld effect \cite{Mei-33}.

In the Ginzburg-Landau theory of superconductivity, it is shown that setting
the integration constant in equation (\ref{E3.4}) to zero is equivalent to
the assumption that the significant spatial variation of the order parameter 
$\psi =\left| \psi \right| e^{i\phi }$ is through the phase $\phi $, and not
the magnitude $\left| \psi \right| $ \cite{Ash-76, Gin-50}. For the density
of Cooper pairs this means that it cannot vary appreciably from its uniform
thermal equilibrium value. For a superfluid neutron star in thermal
equilibrium, where the pairs can flow, but not accumulate or be destroyed,
this should be the case.

To obtain the London penetration depth, we take the static part of the right
equation of (\ref{E2.4})

\begin{equation}
{\bf \nabla \times H}=-\frac{16\pi G}c\;{\bf j}  \label{E3.6}
\end{equation}
and take the curl of it. Using the second London equation (\ref{E3.5}) we
arrive at

\begin{equation}
\nabla ^2{\bf H}=16\pi \frac{G\rho _0}{c^2}\,{\bf H,}  \label{E3.7}
\end{equation}
where we immediately identify the gravitational London penetration depth

\begin{equation}
\Lambda _L=\left( \frac{c^2}{16\pi G\rho _0}\right) ^{1/2}=5.177\times
10^{12}\;\rho _0^{-1/2},  \label{E3.8}
\end{equation}
where the density $\rho _0$ is measured in ${\rm kg/m}^3$ and $\Lambda _L$
is given in meters.

\begin{center}
\section{Coherence Length}
\end{center}

For our derivation of the London equations to be valid we must require that
the London penetration depth $\Lambda _L$ is much larger than the coherence
length $\xi _0$ \cite{deG-66}: 
\begin{equation}
\Lambda _L\gg \xi _0,  \label{E4.1}
\end{equation}
where the coherence length is defined through 
\begin{equation}
\xi _0=\frac{\hbar v_F}{\pi \Delta },  \label{E4.2}
\end{equation}
with $v_F$ the Fermi velocity and $2\Delta $ the pairing energy or energy
gap. We are only interested in a rough estimate of the upper limit of $\xi
_0 $. An upper limit for the Fermi velocity obviously is the speed of light.
If the pairing is caused by the strong nuclear force, as we assumed, then a
lower limit on $\Delta $ is a few hundred thousand keV \cite{Cla-92}. With
these numbers we find for the coherence length 
\begin{equation}
\xi _0\leq 10^{-13}{\rm m},  \label{E4.3}
\end{equation}
which is larger than the separation between two nucleons, but much smaller
than the London penetration depth. Therefore, our derivation of the London
equations was justified in the given approximation.

\begin{center}
\section{Discussion}
\end{center}

We notice that for regular densities occurring in the universe the
gravitational London penetration depth will be a very large number. However,
interesting enough, for neutron stars with densities of about $\rho
_{NS}\cong 2\times 10^{17}{\rm kg/m}^3$, we obtain a London length of $12%
{\rm km}$, which is slightly smaller than the radius of a neutron star with
that density. The London length is inversely proportional to the density,
meaning with increasing density the London penetration depth decreases.

For neutron stars this means that the onset of a gravitational Meissner
effect is predicted. In essence it implies that the gravitational magnetic
field caused by the huge angular momentum of the neutron star, will be
expelled from the center of the neutron star. This could be accomplished
through induced matter-supercurrents in the outer layers of the neutron
stars, creating counter-magnetic fields to expel the gravitational magnetic
field from its interior, in analogy to the electromagnetic case.

A more detailed description is not possible in the approximation used, since
the post-Newtonian approximation is only a rough approximation if used for
neutron stars. We are currently working on the second order post-Newtonian
approximation and numerical simulations \cite{Lan-96}.

\begin{center}
\section{Conclusion}
\end{center}

More significant than the expulsion of the gravitational magnetic field,
however, may be the fact that the Meissner effect is a macroscopic quantum
effect, and in essence, the {\it gravitational} Meissner effect is a
macroscopic quantum {\it gravity} effect. There is experimental evidence
that neutron stars are superfluid as well as superconducting. This indicates
that macroscopic quantum effects do play a significant role for neutron
stars. Furthermore, neutron stars are the strongest gravitational objects
observed with certainty. Under the assumptions used here, we find that
macroscopic quantum gravity effects may play a significant role for neutron
stars.

Since the work of Oppenheimer and Volkoff \cite{Opp-39}, we know that
neutron stars will be stable only up to a certain point. After that the
gravitational pressure overcomes the pressure of the neutron Fermi gas and
continued collapse is predicted. This leads to increasing densities, and in
our model to decreasing London penetration lengths, indicating that
gravitational quantum effects might become more and more important.
Therefore, we would like to emphasize that the prediction of Oppenheimer and
Volkoff is classical in nature, and our work indicates, that a quantum
gravitational treatment may lead to new results.

\begin{center}
\section{Acknowledgment}
\end{center}

I would like to thank Professor V.G.J. Rodgers for spawning my interest in
this topic and for his insightful guidance throughout this work.

\end{document}